# Being at Rest

Douglas M. Snyder

Los Angeles, California[1]

When I was a boy, I had a favorite ride at the amusement park in Santa Monica. I would walk into a large cylinder along with other riders, and we would all stand with our backs against the inside wall of the cylinder. The cylinder would begin to rotate around its axis, and the speed of its rotation would increase until it was rotating very quickly. Then, the floor upon which we were standing would drop away.

Amazingly, we did not fall. In fact, we were glued to the cylinder wall. I felt glued to the wall by a force that was pushing me flat against it and the direction of which was along a line through me and which emanated from the center of the cylinder. I was restrained by the cylindrical wall itself, the counterforce being experienced as equal to the force pushing me outward. What held me up against the force of gravity was the force due to friction resulting from my body pressing against the wall of the cylinder.

For an observer located outside the ride and at rest, the analysis of the motion would be much different. This observer can be considered to be in an inertial reference frame. This observer's analysis would be that the direction of my velocity is always tangent to the rotating cylinder. My velocity would be like a piece of a spinning top that breaks off at the edge of the top. Like the rotating cylinder, the magnitude of the velocity is constant. But, unlike the piece of the spinning top, the direction of my motion is constantly altered due to a force applied to me by the cylinder wall that is responsible for my continual acceleration toward the cylindrical axis (i.e., a centripetal force).[(1)]

In this analysis of the dynamics of my motion, the frictional force due to my body's pressing against the cylinder is also responsible for keeping me from falling. This observer at rest in an inertial reference frame would essentially maintain an explanation of the dynamics of my motion in accordance with Newtonian mechanics. On the other hand, my experience of the dynamics of my motion is similar to analyses of rotational motion maintained by Descartes, where a real outward directed force, for example, is considered responsible for the climbing of water up the inside of a rotating bucket.[(2)]

In the Newtonian analysis of rotational motion, this outward directed, centrifugal force is not essential. Only a centripetal force is required. A complete and accurate understanding of my motion on the amusement park ride is achieved without centrifugal force. Indeed, Newton's analysis of rotational motion, in conjunction with Newton's law of gravitation, is responsible for remarkably accurate predictions concerning the motions of the planets around the sun. The question arises as to why I experienced this outward directed force when it appears that a complete, accurate, and economical explanation does not require it. Is this experience just an insignificant oddity, or can it be seen to reveal something basic about the nature of the physical world and its relation to human experience?

### THE SIGNIFICANCE OF THE OBSERVER'S BEING AT REST IN A REFERENCE FRAME

It is unlikely that this experience of what is called centrifugal force is insignificant. The experience in situations like the amusement park ride is too powerful, and these situations occur too frequently. What then can it tell us about the nature of the physical world and an individual's relation to it? The answer can be found in that which distinguishes between my experience of centrifugal force and the analysis of the observer who is stationary on the ground.

This distinguishing feature concerns the different reference frames of the respective observers. That is, my reference frame on the ride is the rotating cylinder and the reference frame of the stationary observer on the ground is an inertial reference frame. But why should these different reference frames be responsible for the differing analyses and different experiences of the situation? For example, why can't I simply consider myself in the stationary reference frame of the observer on the ground and thus get rid of the experience of centrifugal force? It might appear that one is dealing only with simply choosing one coordinate scheme rather than another from which to describe my motion. And there are certainly transformation rules for transforming the description of motion in one reference frame into the description in another.

The reason for the difficulty in achieving this change of reference frame is the natural inclination to see oneself as an observer at rest in his frame of reference. It is this characteristic of experience that is the basis for distinguishing the experiences and analyses of the different observers of my ride at the amusement park. As implied in the above description of my experience on the ride, the outward directed, centrifugal force is needed by me, or another observer in similar circumstances, to counterbalance the applied

Being at Restcentripetal force. My experience of the counterbalancing forces is associated with my experience of being at rest.

A fundamental aspect of the measurement of motion in the physical world is that the observer involved in this measurement is at rest in some frame of reference. It is a point that is generally not emphasized while the importance of a reference frame in determining the motion of physical objects may be. The reference frame, in part an invention of the observer using it, can itself be problematic. There is no necessity that a particular observer adopt a particular reference frame. Yet, without choosing a particular reference frame, the observer's measuring instruments are essentially useless.

But even more fundamental, from a psychological perspective, is the idea of rest for the observer because this rest is completely beyond physical analysis. This rest is not the rest referred to in Newton's first law of motion. This being at rest for an observer in a reference frame is distinct from the motion or rest of physical objects observed by this individual. If someone else, for example, sees me rotating on the amusement park ride, I nonetheless as an observer am at rest in my reference frame. For this other individual, the amusement park ride is rotating at the same speed as I rotate. Even if I am aware that I, as my body, am rotating at a constant speed, as an observer I consider myself at rest in a reference frame. I am fundamentally at rest even though I may be cognitively convinced that I am in motion.

This last point bears repeating. There is nothing in the physical world that serves as the basis for explaining the observer's being at rest for himself or herself in a reference frame. The observer's being at rest in a reference frame is a psychological characteristic, not an explicitly physical one. The observer's being at rest in a reference frame is a characteristic that follows the psyche, not the material physical world. It does not, for example, follow the physical object associated with the observer's reference frame. There is nothing particularly unique about this physical object that distinguishes it from other physical objects, except that for the observer it is not moving relative to him. Its significance in terms of its, and the associated reference frame's, being at rest is that it is not in motion relative to the observer in the reference frame. If the object to which the observer's reference frame is associated is put into motion relative to him, it is no longer associated with his reference frame as an anchor, so to speak, for this reference frame. Another suitable object is needed. Wherever I am, I am fundamentally at rest, even if I conclude that my body is moving.[2]

- 3 -

# Being at Rest

*Special and General Relativity*

The significance of the observer's being at rest for himself or herself in a reference frame is implied in the theories of special and general relativity. In special relativity, the relativity of simultaneity is the basis for the other results of special relativity, and it depends on observers at rest for themselves in inertial reference frames travelling in uniform rectilinear motion relative to one another. It is from their inertial reference frames, in which the observers are at rest, that the measurements of time can be taken to determine the relativity of simultaneity.[3,4]

In general relativity, one statement of the principle of equivalence is that a uniformly accelerating reference frame is equivalent with regard to the description of motion to an inertial reference frame in a gravitational field of uniform intensity.[5] It is this equivalence that Einstein indicated is the meaning behind the equivalence of inertial mass in Newton's laws of motion and gravitational mass in Newton's law of gravitation.[6] The principle of equivalence underlies the general principle of relativity that states that the laws of physics apply to reference frames whether or not they are inertial ones. In outlining the essentials of the general theory of relativity, Einstein discussed a situation like that of the amusement park ride and posited the importance of the observer's being at rest in the description of physical phenomena. He wrote:

> Let us consider a space-time domain in which no gravitational field exists relative to a reference-body $K$ whose state of motion has been suitably chosen. $K$ is then a Galileian reference-body as regards the domain considered, and the results of the special theory of relativity hold relative to $K$. Let us suppose the same domain referred to a second body of reference $K'$, which is rotating uniformly with respect to $K$. In order to fix our ideas, we shall imagine $K'$ to be in the form of a plane circular disc, which rotates uniformly in its own plane about its centre. An observer who is sitting eccentrically on the disc $K'$ is sensible of a force which acts outwards in a radial direction, and which would be interpreted as an effect of inertial (centrifugal force) by an observer who was at rest with respect to the original reference-body $K$. But the observer on the disc may regard his disc as a reference-body which is "at rest"; on the basis of the general principle of relativity he is justified in doing this. This force acting on himself, and in fact on all other bodies which are

> at rest relative to the disc, he regards as the effect of a gravitational field....he is quite in the right when he believes that a general law of gravitation can be formulated–a law which not only explains the motion of the stars correctly, but also the field of force experienced by himself.[7]

In the above quote, the observer on the disc considers his reference body (i.e., the disc), and thus this body's associated reference frame, at rest. The essential characteristic of the observer's reference frame is that he or she is at rest in it. Thus, in the above example, rather than considering himself accelerating, the observer on the disc considers himself stationary in a gravitational field. If the disc moved differently than did the observer, the observer would need to choose a different reference body that would be associated with his reference frame, one that was moving in the same manner as himself. When Einstein focused on the rotating disc being a reference body at rest for the observer sitting eccentrically on the disc, he essentially was concerned with this observer's being at rest in a reference frame for which the disc was the associated reference body.

CONCLUSION

Exploring the experience of centrifugal force has brought to the fore the natural inclination to see oneself at rest in his or her frame of reference. The discrepancy between the experience of centrifugal force and its lack of fundamental significance in Newton's laws of motion points toward the importance of the observer as a subject in his or her own reference frame, specifically the observer's experience of being at rest in this reference frame. In addition, the observer's being at rest for himself or herself is central to special and general relativity. As discussed, there is nothing in the physical world that serves as the basis for explaining the observer's being at rest in a reference frame. The observer's being fundamentally at rest as a subject allows for the measurement of motion in the physical world. Thus, the measurement of motion in the physical world depends on a feature of experience that cannot be explained in terms of events in the physical world.

ENDNOTES

[1] Email address: dsnyder@earthlink.net.

[2] An anecdote demonstrates the very basic nature of a person's experience of being at rest in a reference frame and, at the same time, the

difficulty in accepting this fact. A few years ago, I was driving in a car with my eight year old niece about two weeks after she had taken her first trip by airplane. As we were riding, my niece asked out of the blue, "Why doesn't it feel like you're moving when you're in an airplane?" She also said, "You get on the airplane at one place and get off the airplane at another, but you don't feel like you've moved anywhere." My niece, like the rest of us, had difficulty squaring her experience that she was not in motion with her having traveled to a far off place. Her experience of not being in motion in the airplane essentially constituted her being at rest in an inertial frame of reference.
## REFERENCES

1. Halliday, D. and Resnick, R. *Physics* (Vol. 1) (3rd ed.). (John Wiley & Sons, 1977), pp. 104-105.

2. Cohen, I.B. Newton's discovery of gravity. *Sci. Amer.*, *244*(3), 166 (1981).

3. Einstein, A. On the electrodynamics of moving bodies. In H. Lorentz, A. Einstein, H. Minkowski, and H. Weyl (Eds.), *The principle of relativity, a collection of original memoirs on the special and general theories of relativity*. (Dover; 1952, orig. work published 1905), pp. 35-69.

4. Einstein, A. *Relativity, the special and the general theory*. (Bonanza; 1961, orig. work published 1917), pp. 17-54.

5. Einstein, A. *The meaning of relativity* (5th ed.) (Princeton University Press, 1956, orig. work published 1922), p. 57.

6. Ref. 5, pp. 57-58.

7. Ref. 4, pp. 79-80.